\documentclass[10pt,conference,a4paper]{elektron}

\usepackage[utf8]{inputenc}
\usepackage[compress]{cite}
\usepackage{graphicx}
\usepackage{float}
\usepackage{amsmath}
\usepackage{multicol}
\usepackage{breqn}
\usepackage{amssymb,amsmath,amsthm} 
\usepackage[spanish,es-noshorthands]{babel}

\def\tablename{Tabla}

\begin{document}
\renewcommand{\tablename}{Tabla} 
\title{Optimizaci\'on del sistema de iluminaci\'on y proyecci\'on para un sensor multiparam\'etrico acusto-\'optico\\ \vspace{0.5cm}
\large Optimization of the illumination and projection system for a multiparametric acousto-optic sensor}

\author{\IEEEauthorblockN{Patricia M. E. Vázquez\IEEEauthorrefmark{2}$^1$ and Germán E. Caro\IEEEauthorrefmark{2}\IEEEauthorrefmark{1}$^2$\vspace{0.2cm}}\IEEEauthorblockA{\IEEEauthorrefmark{2}\emph{Universidad de Buenos Aires, Facultad de Ingenier\'ia, Departamento de F\'isica, GLOmAe.}\\\emph{Paseo Col\'on 850, C1063ACV, Buenos Aires, Argentina.}}\IEEEauthorblockA{$^1$\texttt{\small{pvazquez@fi.uba.ar}}}\IEEEauthorblockA{\IEEEauthorrefmark{1}\emph{Consejo Nacional de Investigaciones Cient\'ificas y T\'ecnicas de Argentina (CONICET).}\\\emph{Godoy Cruz 2290, C1425FQB, Buenos Aires, Argentina.}\\$^2$\texttt{\small{gcaro@fi.uba.ar}}}}



\maketitle

\begin{abstract}

En trabajos anteriores presentamos un novedoso sensor multiparamétrico que mide simultáneamente el índice de refracción y la velocidad del sonido en un líquido aprovechando el efecto acusto-óptico. El sensor requiere un sistema de iluminación que expanda el haz láser de manera tal que interactúe eficazmente con el líquido bajo estudio. Por otra parte, también es necesario un sistema de proyección a fin de capturar adecuadamente el patrón de difracción a la salida de la celda. En este sentido, en este artículo, presentamos la optimización del sistema de iluminación y proyección del sensor utilizando la herramienta computacional de diseño óptico Zemax OpticStudio. El rango considerado de índices de refracción de las muestras líquidas está comprendido entre 1,33 y 1,51. Como resultado se obtuvo una correcta expansión del haz incidente sobre la celda, y el sistema de proyección logra una imagen con muy pocas aberraciones y una separación angular adecuada entre los máximos del patrón de difracción. A su vez, las dimensiones resultantes del sensor luego de la optimización del sistema de iluminación y proyección permiten afirmar que es compacto y portable.
\\
\\
Palabras clave: diseño de sistema óptico; optimización; aberraciones.  
\\
\\
$~~$\emph{Abstract---} 
In previous works we presented a novel multiparametric sensor to simultaneously measure the refractive index and the speed of sound in a liquid by means of the acousto-optic effect. The sensor requires an illumination system that expands the laser beam so that it interacts effectively with the liquid under study. Also, a projection system is necessary in order to adequately capture the diffraction pattern at the cell exit. In this article, we present the optimization of the sensor illumination and projection system using the optical design computational tool Zemax OpticStudio. The considered range of refractive indices of the liquid samples is between 1.33 and 1.51. The results show a correct expansion of the incident beam on the cell, and the projection system achieves an image with very few aberrations and an adequate angular separation between the maxima of the diffraction pattern. Based on the resulting dimensions of the sensor after optimizing the illumination and projection system, we confirm that it is compact and portable.
\end{abstract}


\IEEEpeerreviewmaketitle

\section{Introducción}
La caracterizaci\'on no destructiva de productos l\'iquidos de biomasa, como los biocombustibles, es de creciente relevancia \cite{Corach2015,Corach2021,Turck2022}. En particular, es importante la caracterización de productos de biomasa para verificar su calidad e identificar contaminantes o adulterantes en el transporte, el almacenamiento y la distribución. En la actualidad, existen técnicas estandarizadas para la caracterización de biocombustibles, como la espectroscopia infrarroja y la cromatografía. Sin embargo, los instrumentos de medición empleados en dichas técnicas no son portables y son de elevado costo. Es de interés emplear nuevas técnicas y estudiar distintas propiedades para la caracterización de líquidos de biomasa \cite{Sorichetti2005,Romano2011,Corach2019,Mandalunis2021,Lichtinger2023}. 

Las características ópticas \cite{Colman2018,Corach2023,Romano2023} y acústicas \cite{Sorichetti2024} de los biocombustibles aportan información sobre propiedades relacionables con la composición y el estado del líquido. Un trabajo previo \cite{Oreglia2019} presenta un sensor multiparam\'etrico en forma de celda en cuña, que aprovecha la interacci\'on luz-materia para medir la velocidad del sonido en el líquido y su \'indice de refracci\'on. Este dispositivo es de bajo costo, no requiere personal especializado para su utilización y es portable, es decir, que con el mismo se pueden realizar mediciones ``in-situ", en campo o en planta.

\begin{figure}[!h]
\centering
\includegraphics[clip,width=0.95\columnwidth]{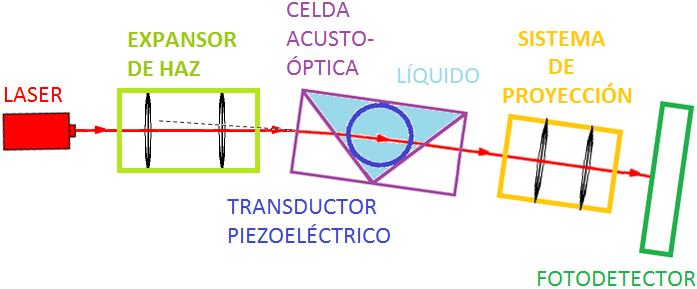}
\caption{Vista superior del sensor multiparamétrico acusto-óptico, en la que se indican el expansor de haz del sistema de iluminación y el sistema de proyección.}
\label{FIG:figura1}
\end{figure}

La celda en cuña para refractometría o celda de Hughes \cite{Hughes1941} está compuesta por dos prismas de un material transparente que forman una cavidad donde se coloca la muestra. En la celda acusto-óptica se dan cuatro refracciones del haz láser de manera tal que, ajustando el ángulo incidente, el haz sale perpendicular a la superficie de salida de la celda (Fig. \ref{FIG:figura1}). 
A partir de la medición del ángulo incidente sobre la celda, se puede determinar el índice de refracción del líquido \cite{Vazquez2020}.

\begin{figure}[!h]
\centering
\includegraphics[clip,width=0.95\columnwidth]{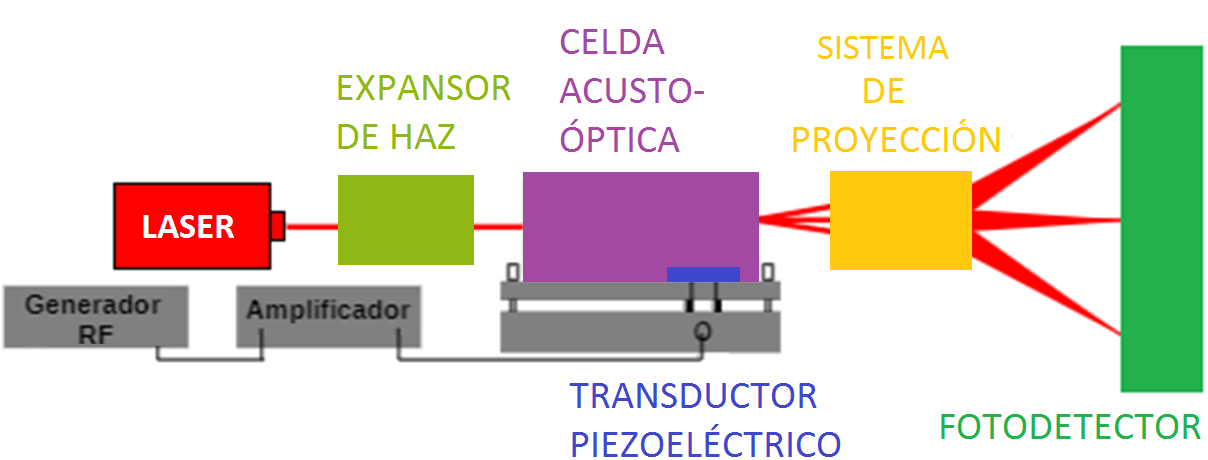}
\caption{Vista lateral del sensor multiparamétrico acusto-óptico, en la que se muestran los elementos para inyectar ultrasonido en el líquido que se encuentra en la celda, y se indican el expansor de haz del sistema de iluminación y el sistema de proyección.}
\label{FIG:figura2}
\end{figure}

En la vista lateral del sensor (Fig. \ref{FIG:figura2}), se muestra que se inyecta ultrasonido en el líquido que se encuentra en la celda a través de un generador de radiofrecuencia (RF) conectado a un amplificador, y luego al transductor piezoeléctrico. 
Dentro de la celda acusto-óptica, el transductor piezoeléctrico origina una onda ultrasónica en el líquido, que a su vez produce variaciones periódicas de su índice de refracción. Cuando incide un haz de luz láser perpendicular a la dirección de propagación del ultrasonido, se da una modulación de fase en el frente de onda del haz. El líquido se comporta como una red de fase, dando lugar a un patrón de difracción (efecto acusto-óptico) \cite{Raman1935}. 
El efecto acusto-óptico es utilizado para diversas aplicaciones, como microscopía e imágenes biomédicas \cite{Duocastella2021} \cite{MestreTora2023}, fotónica integrada en niobato de litio \cite{Pan2023}, o giróscopos de onda progresiva \cite{Tian2022}. 

La velocidad del sonido se determina a partir del patrón de difracción \cite{Oreglia2019}. En el fotodetector se proyecta el diagrama de difracción originado en la celda acusto-óptica (Fig. 2). A través del fotodetector del sensor acusto-óptico, se miden las posiciones de los máximos del patrón de difracción proyectados y las distancias entre los mismos. Por lo tanto, una configuración adecuada para el diseño del detector óptico es un arreglo de fotodiodos discretos \cite{Sorichetti2000} \cite{Vazquez2018}. En una etapa del diseño del sensor posterior a la optimización del sistema de iluminación y proyección, en la que se diseñará e implementará el fotodetector, será muy importante tener en cuenta las especificaciones de relación señal a ruido, ancho de banda y margen de fase \cite{Vazquez2021}.

Los sistemas ópticos reales, al enfocarlos, dan imágenes distribuidas en el espacio en lugar de un punto. Estas propiedades, llamadas aberraciones, dan lugar a imágenes borrosas o distorsionadas, que dependen de la naturaleza de la aberración. Las aberraciones pueden ser estudiadas como desvíos de la aproximación paraxial, y deben ser tenidas en cuenta y corregidas antes de fabricar o implementar un sistema óptico \cite{gross2005handbook}. Una de las aberraciones más comunes es la aberración esférica, donde los rayos con distinto ángulo tienen el foco en el mismo eje pero distinta posición, y surgen como consecuencia de utilizar lentes con caras esféricas. 

La alta complejidad del fenómeno de las aberraciones hace difícil un cálculo analítico o numérico manual, y dificulta la correcta elección de las piezas ópticas necesarias. Por este motivo, existen numerosas herramientas computacionales de diseño óptico que, utilizando óptica geométrica no aproximada, permiten no sólo ver el efecto de estas aberraciones, sino también encontrar la combinación de parámetros característicos de cada elemento óptico que las minimice. Esto lo realizan internamente creando una función que incluya los distintos criterios de optimización que se deseen conseguir, que pueda ser calculada de manera numérica.

\begin{figure*}[!h]
\centering
\includegraphics[width=0.8\linewidth]{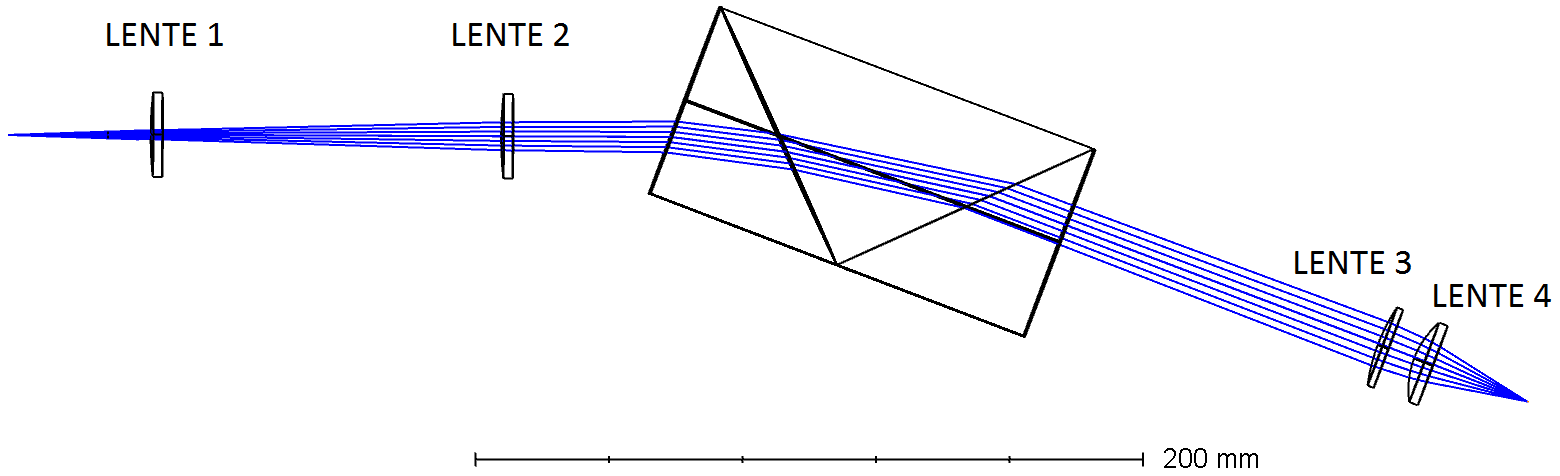}
\caption{Trazado de rayos sobre el esquema experimental del sensor con el sistema de iluminación (lentes 1 y 2) y proyección (lentes 3 y 4) optimizado.}
\label{fig:133diag}  
\end{figure*}

En este trabajo, presentamos la optimización del  sistema de iluminación y proyección del sensor acusto-óptico para la medición simultánea y no destructiva del índice de refracción y de la velocidad del sonido en líquidos (Fig. \ref{FIG:figura1} y \ref{FIG:figura2}). Respecto al sistema de iluminación, es importante expandir el haz de luz a la entrada de la celda acusto-óptica para que interactúe eficazmente con el líquido bajo estudio. A su vez, es necesario utilizar un sistema de proyección a la salida de la celda acusto-óptica para lograr una separación angular adecuada entre los máximos del patrón de difracción (Fig. \ref{FIG:figura2}). Realizamos la optimización de los parámetros de las lentes que forman el expansor de haz del sistema de iluminación y el sistema de proyección utilizando la herramienta computacional para el diseño de sistemas ópticos, Zemax OpticStudio \cite{Zemax}. 

El artículo se estructura de la manera que detallamos a continuación. En la Sección II, presentamos el esquema experimental y el procedimiento de la optimización del sistema de iluminación y proyección del sensor. Luego, en la Sección III, mostramos los resultados de la optimización. Finalmente, informamos las conclusiones y las perspectivas de este trabajo, en la Sección IV.

\section{Esquema experimental y procedimiento de la optimización}

El sensor multiparamétrico acusto-óptico (Figs. \ref{FIG:figura1} y \ref{FIG:figura2}) está compuesto por un láser de He-Ne, de $632,8$ nm; un expansor de haz (sistema de iluminación); 
una celda acusto-óptica de $3$ cm de lado y $6$ cm de largo, con un transductor piezoeléctrico en el fondo de la cavidad y prismas de vidrio de índice de refracción $1,51$; un amplificador; un generador RF; un sistema de proyección; y un fotodetector. Realizamos la optimización del sistema de iluminación,  formado por las lentes 1 y 2, y del sistema de proyección, compuesto por las lentes 3 y 4 (Fig. \ref{fig:133diag}). 

El sensor está diseñado para caracterizar líquidos de índices de refracción $1,33$, $1,40$ y $1,51$, y los comprendidos entre estos valores. En particular, el índice de refracción $1,40$ es de interés por ser próximo al de muchos productos líquidos de biomasa. El índice de refracción del agua es aproximadamente 1,33. No consideramos valores superiores al índice de refracción del vidrio (1,51) visto que la mayoría de los fluidos de interés biotecnológico tienen índice de refracción por debajo de ese valor. 

Otras especificaciones de la optimización son que el sensor sea compacto, que el sistema de iluminación expanda el haz en la mayor medida posible pero considerando como diámetro máximo a la entrada de la celda $7,500$ mm, y que el sistema de proyección permita alcanzar un foco adecuado sobre el fotodetector del sensor. Cabe destacar que para el sistema de iluminación y para el sistema de proyección hay requerimientos distintos. Esto nos permite dividir el problema en dos etapas independientes. Por otra parte, para facilitar el proceso de fabricación del sensor acusto-óptico, optimizamos el sistema utilizando sólo lentes comerciales.



\begin{figure}[!h]\centering
	\includegraphics[clip,width=0.95\columnwidth]{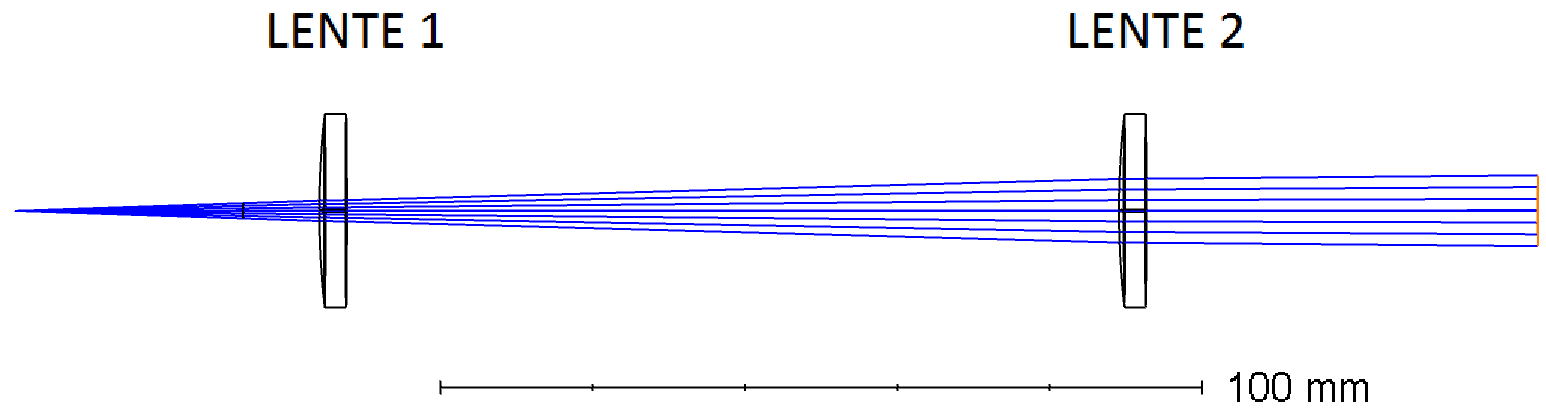}
	\caption{Expansor de haz del sistema de iluminación optimizado.} 
	\label{fig:133esquemaexpansor}
\end{figure}

En primer lugar, diseñamos el expansor de haz del sistema de iluminación separado del resto de las partes del sensor (Fig. \ref{fig:133esquemaexpansor}). 
El objetivo es lograr una iluminación uniforme en la superficie de la celda. A través de la herramienta computacional, optimizamos los parámetros de las lentes 1 y 2, es decir, sus distancias focales y sus posiciones, para obtener en un espacio compacto un haz expandido adecuadamente pero de diámetro menor a $7,500$ mm, y con pocas aberraciones. 

Luego de optimizado el sistema de iluminación, diseñamos el sistema de proyección. Dejamos fijos los valores previamente hallados para el sistema de iluminación, y tomamos como parámetros a optimizar las distancias focales y las posiciones de las lentes 3 y 4 del sistema de proyección (Fig. \ref{fig:133diag}). El objetivo en este caso es obtener el mejor foco posible, usando como criterio minimizar tanto el radio cuadrático medio del haz proyectado en el fotodetector, 
como las aberraciones presentes. 




En la siguiente sección, presentamos los resultados que obtuvimos de la optimización del sistema de iluminación y proyección del sensor multiparamétrico acusto-óptico.

\section{Resultados de la optimización}

\begin{figure}[!h]\centering
	\includegraphics[clip,width=0.5\columnwidth]{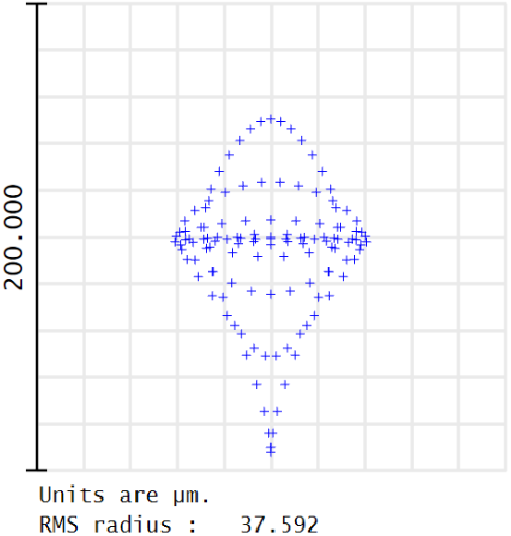}
	\caption{Diagrama de puntos (\textit{spot diagram}) del sistema de iluminación y proyección optimizado, para el caso en el que en la celda hay un líquido de índice de refracción $1,33$.}
	\label{fig:133spot}
\end{figure}

\begin{figure}[!h]\centering
	\includegraphics[clip,width=1\columnwidth]{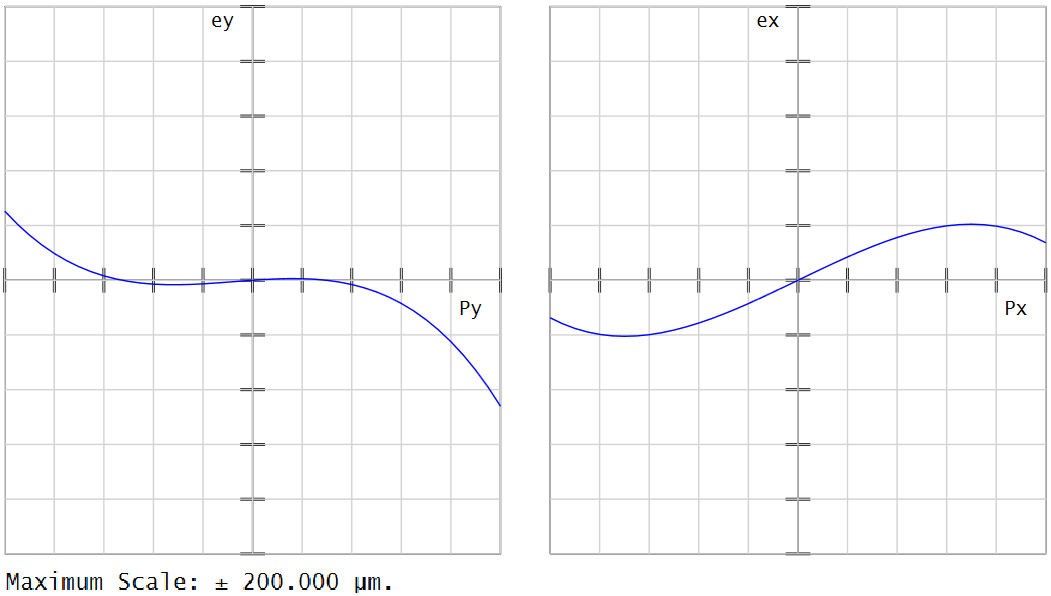}
	\caption{Perfil transversal del haz de rayos (\textit{transverse ray fan plot}) del sistema de iluminación y proyección optimizado.}
	\label{fig:133ray}
\end{figure}

\begin{figure}[!h]\centering
	\includegraphics[clip,width=1\columnwidth]{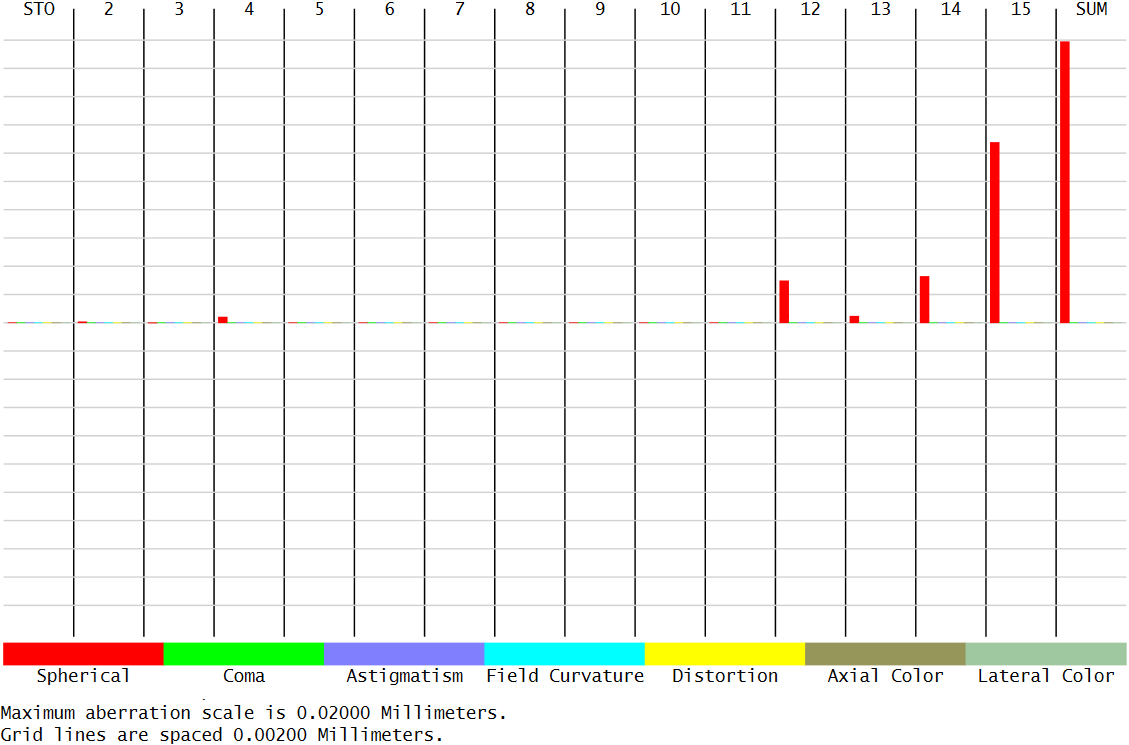}
	\caption{Diagrama de Seidel del sistema de iluminación y proyección optimizado para el caso en el que en la celda hay un líquido de índice de refracción $1,33$.}
	\label{fig:133seidel}
\end{figure}

La herramienta computacional Zemax OpticStudio \cite{Zemax} cuenta con distintas herramientas para observar el desempeño del sistema óptico que estamos optimizando. Podemos observar el trazado de rayos sobre el esquema del sistema óptico (Fig. \ref{fig:133diag}), el diagrama de puntos (\textit{spot diagram}, Fig. \ref{fig:133spot}), el perfil transversal del haz de rayos (\textit{transverse ray fan plot}, Fig. \ref{fig:133ray}) y el diagrama de Seidel (Fig. \ref{fig:133seidel}), entre otras opciones. Todas estas herramientas permiten analizar la calidad del sistema óptico y sus aberraciones.


El diagrama de Seidel se utiliza para analizar y visualizar las aberraciones en un sistema óptico. Este tipo de diagrama muestra la contribución de los diferentes tipos de aberraciones, como la esférica, coma, astigmatismo, curvatura de campo y distorsión (a cada una le corresponde un color distinto, como se indica en la Fig. \ref{fig:133seidel}). Para leer el diagrama de Seidel, se deben observar las barras, donde la desviación respecto de la línea base indica la magnitud, y el color, el tipo de aberración presente. Una barra más corta y, por ende, menos desviada de la línea base, sugiere una menor aberración y, por ende, un mejor rendimiento óptico. Los números que se observan en la parte superior de la Fig. \ref{fig:133seidel} corresponden al orden de los coeficientes de Seidel \cite{gross2005handbook}.

Realizamos los múltiples pasos de la optimización del sistema de iluminación y proyección del sensor multiparamétrico acusto-óptico para el caso en el que en la cavidad de la celda en cuña hay un líquido de índice de refracción $1,33$. Este el caso en el que ángulo de incidencia del  haz sobre la superficie de entrada de la celda (medido respecto a la recta perpendicular a la superficie de la celda, indicada con una línea punteada en la Fig. \ref{FIG:figura1}) es máximo, como se muestra en el esquema experimental de la Fig. \ref{fig:133diag}.
Optimizamos el diseño del sistema para este caso extremo debido a que, si funciona correctamente para dicho caso, también es adecuado para los restantes casos (líquidos de índices de refracción $1,40$ y $1,51$), en los que el haz incide en la celda con ángulos menores. Partimos del peor caso para tener una única configuración experimental que sea adecuada para todo el rango de interés. Cabe destacar que es importante, al finalizar el diseño optimizado, verificar el correcto desempeño del sistema óptico cuando en la celda hay líquidos de los otros dos índices de refracción considerados para analizar la envolvente de diseño.


En una primera instancia, dejamos a la herramienta computacional de diseño óptico variar libremente los parámetros que caracterizan a las lentes, y como resultado no obtuvimos lentes comerciales. Si bien es posible adquirir estas lentes construidas a medida por distintos fabricantes, su costo es elevado y poco práctico. Por este motivo, procedimos a utilizar el catálogo de lentes comerciales de Zemax OpticStudio \cite{Zemax}, provisto por cada fabricante, y buscamos las más cercanas a las de nuestro sistema \cite{Lente1}. Probamos mediante la herramienta computacional el sistema con los valores de las cuatro lentes comerciales (Tabla \ref{TAB:tab1}), como detallamos a continuación, y los resultados no se vieron afectados. 

Las lentes comerciales del sistema de iluminación y proyección y los valores de sus parámetros resultado del proceso de optimización se muestran en la Tabla \ref{TAB:tab1}.
\begin{table}[H]
\centering
\caption{Lentes comerciales resultado de la optimización.}
\begin{tabular}{|p{0.058\columnwidth}|p{0.26\columnwidth}|p{0.2\columnwidth}|p{0.291\columnwidth}|} \hline
\textbf{Lente} & \textbf{Marca} & \textbf{Modelo} & \textbf{Distancia focal (mm)} \\ \hline
1 & DAHENG OPTICS & GCL-010165A & 222,60 \\ \hline
2 & DAHENG OPTICS & GCL-010165A & 222,60 \\ \hline
3 & DAHENG OPTICS & GCL-010111B & 101,20 \\ \hline
4 & DAHENG OPTICS & GCL-010107 & 50,80 \\ \hline
\end{tabular}
\label{TAB:tab1}
\end{table}
Las posiciones de los elementos ópticos del sistema, que también son resultado de la optimización, se indican a través de las distancias entre los mismos, en la Tabla \ref{TAB:tab2}. 
\begin{table}[H]
\centering
\caption{Posiciones optimizadas de los elementos ópticos.}
\begin{tabular}{|p{0.45\columnwidth}|p{0.22\columnwidth}|} \hline
\textbf{Elementos ópticos} & \textbf{Distancia (mm)} \\ \hline
  Objeto (has gaussiano) - Lente 1 & 40  \\ \hline 
 Lente 1 - Lente 2 & 102  \\ \hline 
  Lente 2 - Celda & 51   \\ \hline 
 Celda - Lente 3 & 100 \\ \hline 
 Lente 3 - Lente 4 & 7\\ \hline 
 Lente 4 - Imagen (Fotodetector) & 30\\ \hline 
\end{tabular}
\label{TAB:tab2}
\end{table}

\begin{figure}[!h]\centering
\includegraphics[clip,width=0.5\columnwidth]{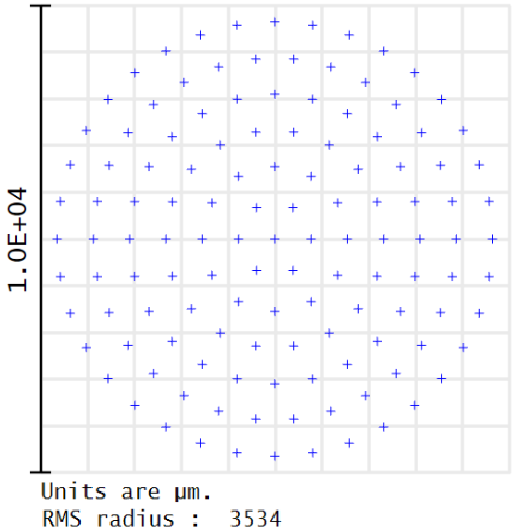}
	\caption{Diagrama de puntos (\textit{spot diagram}) del sistema de iluminación (expansor de haz) optimizado.}
	\label{fig:133spotexpansor}
\end{figure}

En la  Fig. \ref{fig:133diag}, se puede ver el esquema del sistema óptico optimizado para el caso en el que en la celda hay un líquido de índice de refracción $1,33$.
Evaluamos por partes el sistema óptico formado por las cuatro lentes comerciales. Comenzamos por el sistema de iluminación, es decir, el expansor de haz 
(lentes 1 y 2, Fig. \ref{fig:133esquemaexpansor}). Verificamos el tamaño del haz a la salida del expansor con las lentes comerciales previamente seleccionadas, y resulta de $3,534$ mm de radio cuadrático medio, como se muestra en el diagrama de puntos de la Fig. \ref{fig:133spotexpansor}. 
Por lo tanto, el diámetro del haz a la entrada de la celda es de $7,068$ mm, y cumple con la especificación, ya que es menor a $7,500$ mm. 

A continuación, 
analizamos el desempeño del sistema de iluminación y proyección completo, y obtuvimos una imagen con muy pocas aberraciones y un radio cuadrático medio de $37,592\ \mu$m (Fig. \ref{fig:133spot}). 
A través del diagrama de puntos observamos que, en la posición en la que se ubica la imagen del sensor multiparamétrico, 
 los rayos no convergen a un punto, como se puede esperar para el caso de un sistema óptico real y no aproximado.
Hay aberraciones presentes debido a que no se forma una imagen circular en el diagrama de puntos. Esto lo podemos verificar en el perfil transversal del haz de rayos (Fig. \ref{fig:133ray}), y también en el diagrama de Seidel (Fig. \ref{fig:133seidel}), que confirman la presencia únicamente de aberraciones esféricas, de alto orden.
El orden y su bajo valor nos indican que son poco significativas, y que el radio cuadrático medio sea del orden de los micrones nos permite concluir que obtuvimos buenos parámetros de diseño.


%

\begin{figure}[!h]\centering
	\includegraphics[clip,width=0.5\columnwidth]{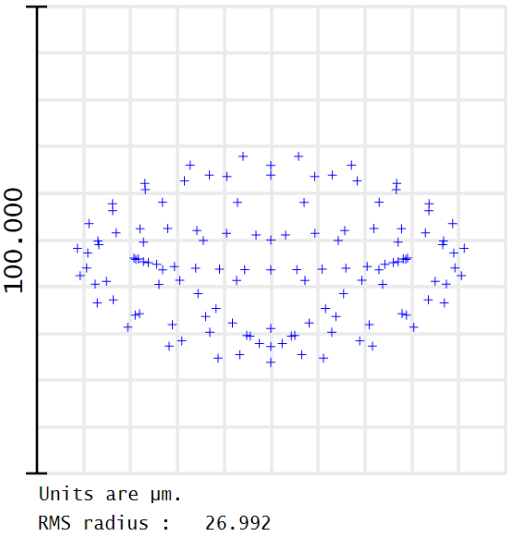}
	\caption{Diagrama de puntos (\textit{spot diagram}) del sistema de iluminación y proyección optimizado para el caso del líquido de índice de refracción $1,33$, con una muestra de índice $1,40$ dentro de la celda. }
	\label{fig:140spot}
\end{figure}

\begin{figure}[!h]\centering
	\includegraphics[clip,width=0.5\columnwidth]{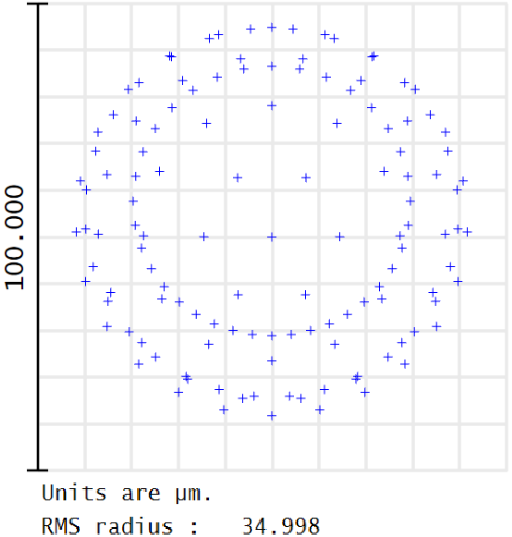}
	\caption{Diagrama de puntos (\textit{spot diagram}) del sistema de iluminación y proyección optimizado para el caso del líquido de índice de refracción $1,33$,
    con una muestra de índice $1,51$ dentro de la celda. }
	\label{fig:151spot}
\end{figure}

Para verificar el correcto funcionamiento del sistema de iluminación y proyección para los restantes índices de refracción especificados de líquidos ($1,40$ y $1,51$), 
modificamos el índice de refración de la muestra a través de la herramienta computacional. A su vez, sabiendo que el ángulo de incidencia en la superficie de entrada de la celda se modifica al colocar un líquido de un índice de refracción distinto en la celda, ajustamos también manualmente el ángulo del eje óptico del conjunto celda + lente 3 + lente 4, para garantizar que el haz salga perpendicular a la superficie de salida de la celda. Cambiamos el índice de refracción de la muestra a $1,40$, y en el diagrama de puntos (Fig. \ref{fig:140spot}), observamos que el radio cuadrático medio es de $26,992\ \mu$m y hay pocas aberraciones. 
En el caso en el que el líquido tiene el mismo índice que los prismas de la celda ($1,51$), vemos que el radio cuadrático medio de la imagen en el fotodetector es de $34,998\ \mu$m, con pocas aberraciones presentes (Fig. \ref{fig:151spot}). 

Es importante destacar, por último, que el tamaño del sensor diseñado a través de la optimización es compacto, de $453$ mm de largo y $117$ mm de ancho. Por lo tanto, logramos una optimización del sistema de iluminación y proyección del sensor multiparamétrico acusto-óptico que verifica todas las especificaciones.

\section{Conclusiones}

En este trabajo presentamos la optimización del sistema de iluminación y proyección de un sensor multiparamétrico acusto-óptico que aprovecha la interacción luz-materia para medir simultáneamente el índice de refracción y la velocidad del sonido en un fluido. Este sensor es de interés para la caracterización no destructiva de líquidos de biomasa, como los biocombustibles. 


A través de la utilización 
de la herramienta computacional de diseño óptico Zemax OpticStudio, optimizamos el sistema de iluminación y proyección del sensor, y analizamos su desempeño. Los resultados indican que el sistema de iluminación expande el haz láser a un diámetro de $7,068$ mm. La configuración optimizada de dos lentes del expansor de haz y dos del sistema de proyección presenta aberraciones esféricas, sin embargo, son predominantemente de alto orden. 
Por lo tanto, el sistema de proyección enfoca el haz en el fotodetector, con un radio cuadrático medio por debajo de $38\ \mu$m, para líquidos de los índices de refracción especificados de $1,33$, $1,40$ y $1,51$. El sensor con el sistema de iluminación y proyección optimizado es compacto, visto que sus dimensiones resultantes son $453$
mm de largo y $117$ mm de ancho. En conclusión, la optimización del sistema de iluminación y proyección del sensor verifica ampliamente las especificaciones. 

Futuras investigaciones se enfocarán en explorar el impacto de la utilización de luz policromática en el sistema y la integración del sistema óptico con el fotodetector en el montaje experimental.




\section*{Agradecimientos}
Este trabajo fue realizado con el aporte de Universidad de Buenos Aires (UBACyT 20020190100275BA y 20020190100032BA), CONICET (11220200102112CO) y ANPCyT (PICT-2020-SERIEA-03741). Tesis doctoral de P. M. E. V\'azquez desarrollada en el marco de la beca Peruilh FIUBA. Tesis de G. E. Caro desarrollada en el marco de la beca interna doctoral CONICET.


%

\end{document}